\documentclass[3p,times,procedia]{elsarticle}
\usepackage{grffile}
\usepackage[hyphens]{url}
\usepackage{graphicx,xcolor}
\usepackage{amsmath,amssymb,commath}
\usepackage{xspace}
 \usepackage{nupha_ecrc}

\volume{00}
\firstpage{1}
\journalname{Nuclear Physics A}
\runauth{C.A.G. Prado et al.}
\jid{nupha}
\jnltitlelogo{Nuclear Physics A}
\newcommand\dabmod{\textsc{dab-m}od\xspace}
\newcommand\ROOT{\textsc{root}\xspace}
\newcommand\pythia{\textsc{pythia8}\xspace}
\newcommand\qcd{\textsc{qcd}\xspace}
\newcommand\fonll{\textsc{fonll}\xspace}
\newcommand\qgp{\textsc{qgp}\xspace}
\newcommand\gflow{\ensuremath{\Gamma_\text{flow}}\xspace}
\newcommand\Dmeson{D$^0$\xspace}

\newcommand\raa{\ensuremath{R_\text{AA}}\xspace}
\newcommand\vusphydro{v-\textsc{usp}hydro\xspace}
\newcommand\mckln{\textsc{mckln}\xspace}
\begin{document}
\begin{frontmatter}
  \dochead{}
  \title{Event-by-event $v_n$ correlations of soft hadrons and heavy mesons
      in heavy ion collisions}
  \author[a]{Caio A.~G.~Prado}
  \author[b]{Jacquelyn Noronha-Hostler}
  \author[a]{Roland Katz}
  \author[a]{Jorge Noronha}
  \author[a]{Marcelo G.~Munhoz}
  \author[a]{Alexandre A.~P.~Suaide}
  \address[a]{Instituto de F\'{i}sica, Universidade de S\~{a}o Paulo, C.P.
      66318, 05315-970 S\~{a}o Paulo, SP, Brazil}
  \address[b]{Department of Physics, University of Houston, Houston TX 77204,
      USA}

  \begin{abstract}
    Combining event-by-event hydrodynamics with heavy quark energy loss we
    compute correlations between the heavy and soft sectors for elliptic and
    triangular flow harmonics $v_2$ and $v_3$ of \Dmeson mesons in PbPb
    collisions at $2.76$ TeV and $5.02$ TeV. Our results indicate that $v_3$
    is strongly influenced by the fragmentation temperature and that it
    builds up later than $v_2$ during the evolution of the system.
  \end{abstract}
\end{frontmatter} 

\section{Introduction}\label{sec:intro}

It is known that final state flow anisotropies are converted from medium
density gradients present in early stages of heavy ion collisions due to the
nearly perfect fluidity property of the Quark-Gluon Plasma (\qgp).
Event-by-event viscous hydrodynamics has been shown to accurately describe
the anisotropic flow coefficients, $v_n$, in the soft limit ($p_T < 2$
GeV)~\cite{Noronha-Hostler:2015uye}.  However, at high $p_T$ the underlying
physical mechanism behind anisotropic flow changes and $v_n$ is driven by
differences in the path length of jets flowing through the
plasma~\cite{Noronha-Hostler:2015uye, Wang:2000fq}, a picture that has been
confirmed by event-by-event jet energy loss combined with viscous
hydrodynamics calculations~\cite{Noronha-Hostler:2016eow}. In this picture,
there is an approximate linear response relation between the high $p_T$ $v_2$
and the initial state eccentricity $\epsilon_2$.

Recent calculations using event shape engineering
techniques~\cite{Schukraft:2012ah,Aad:2015lwa} has shown that heavy flavor
meson azimuthal anisotropy at high $p_T$ are linearly correlated with the
anisotropy in the soft sector~\cite{Prado:2016szr}.  Following these
calculations, in this proceeding, we further investigate the correlations
between \Dmeson mesons with $p_T \gtrsim 10$ GeV to all charged particles in
the soft sector for PbPb at $\sqrt{s} = 2.76$ TeV and $\sqrt{s} = 5.02$ TeV
collisions.  This is done by combining a heavy quark energy loss model with
event-by-event viscous hydrodynamic backgrounds, which allows for computing
the nuclear modification factor, \raa, and the corresponding flow
coefficients $v_2$ and $v_3$.

\section{Development of the simulation}

In order to study the evolution of the heavy quarks inside the \qgp we
developed the so-called \dabmod~\cite{Prado:2016szr}, a modular Monte Carlo
simulation program written in C++, using \ROOT~\cite{Brun:1997pa} and
\pythia~\cite{Sjostrand:2007gs} libraries. The modular characteristic of the
program allows for one to select different energy loss models, medium
backgrounds or hadronization processes while studying the evolution of the
system. In the simulation, bottom and charm quarks are sampled within the
transverse plane at midrapidity of the \qgp medium with their initial
momentum given by p\qcd calculation using
\fonll~\cite{Cacciari:1998it,Cacciari:2001td}. Each sampled heavy quark
travels along the transverse plane with a velocity $v$ and a constant
direction $\varphi_\text{quark}$. We implement a simple parametrization of
the energy loss per unit length given as:
\begin{equation}
  \od{E}{x}(T,v) = -f(T,v) \gflow,
\end{equation}
where $T$ is the local temperature, $\gflow = \gamma[1 -
v_\text{flow}\cos(\varphi_\text{quark} - \varphi_\text{flow})]$ (with $\gamma
= 1/\sqrt{1-v_\text{flow}^2}$) is the flow factor with $\varphi_\text{flow}$
the local azimuthal angle of the underlying flow.

In this work we consider $f(T,v) = \alpha$, inspired by the study performed
in Ref.~\cite{Das:2015ana}, which showed that a non decreasing drag
coefficient near the phase transition is favored for a simultaneous
description of heavy flavor $\raa(p_T)$ and $v_2(p_T)$.  The free parameter
$\alpha$ in the energy loss expression is fixed by matching the \Dmeson \raa
computed by \dabmod to data for $p_T \sim 10$ GeV in the central collisions.

We use the \vusphydro event-by-event relativistic viscous hydrodynamical
model~\cite{Noronha-Hostler:2013gga,Noronha-Hostler:2014dqa,Noronha-Hostler:2015coa}
for the temperature and flow profiles of the medium.  For initial conditions,
\mckln \cite{Drescher:2006ca} is used with $\eta/s = 0.11$ and an initial
time $\tau_0 = 0.6$ fm, which leads to a good description of experimental
data for the flow harmonics at low $p_T$.  Currently, no coalescence is
implemented in the code and hadronization of the heavy quarks is assumed to
occur when the local temperature reaches a chosen temperature $T_d$, at which
fragmentation~\cite{Peterson:1982ak} is performed.  Also, no effect on the
medium from the traversing heavy quarks is considered during this calculation
and the heavy quarks are treated as probes.

The event-by-event analysis uses a couple of thousand hydro events in each
centrality bin.  Heavy quarks are oversampled for each event. That
allows us to compute the nuclear modification factor $\raa^q(p_T,\varphi)$,
for a given heavy quark flavor $q$ or heavy meson from $q$, and its
corresponding flow coefficients $v_n^q$.  The reason for the oversampling is
to give a sufficient probability to find $v_n^q(p_T)$ in a hydro event with a
certain $v_n$ in the soft sector. From the flow coefficients in the hydro
events we compute the multi-particle
cumulants~\cite{Luzum:2012da,Luzum:2013yya} following the procedure performed
in Refs.~\cite{Betz:2016ayq} using multiplicity weighting and centrality
rebinning.

\section{Results}\label{sec:resul}

\begin{figure}[b]
  \centering
  \includegraphics{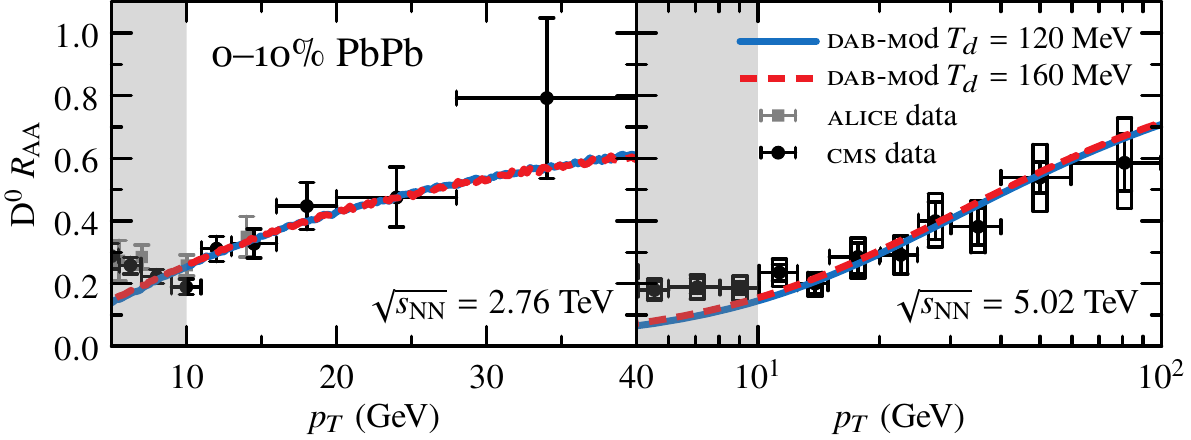}
  \caption{(Color online) \Dmeson nuclear modification factor \raa computed
    by \dabmod for $\sqrt{s} = 2.76$ TeV (left) and $\sqrt{s} = 5.02$ TeV
    (right) PbPb 0--10\% central collisions.  Gray area indicates $p_T$
    region where effects of coalescence may be significant.  Experimental
    data from Refs.~\cite{Adam:2015sza, CMS-PAS-HIN-15-005,
    CMS-PAS-HIN-16-001}.}
  \label{fig:raa_meson}
\end{figure}

\begin{figure}[t]
  \centering
  \includegraphics{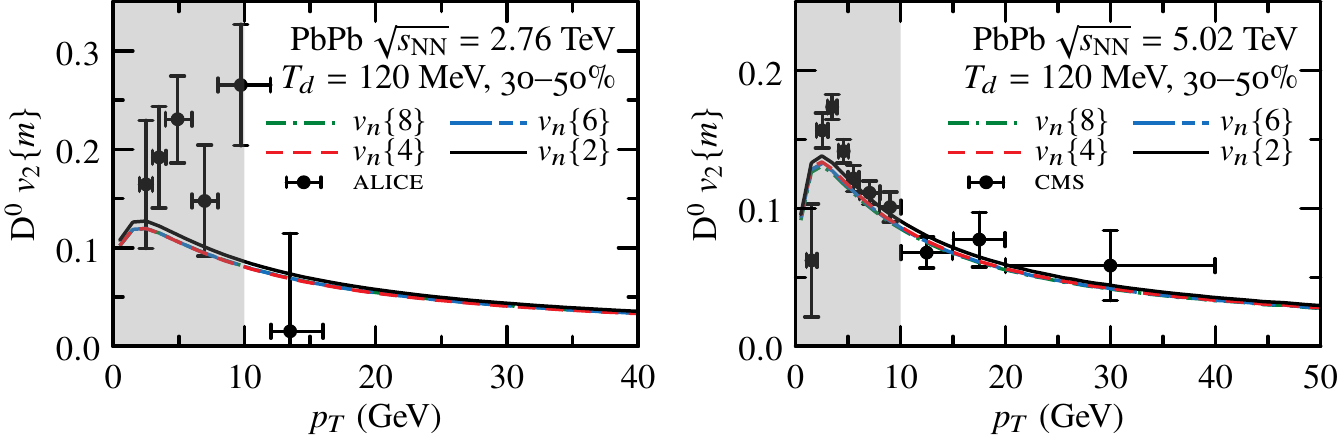}
  \caption{(Color online) Multi-particle cumulants for elliptic flow
    $v_2\{m\}$ of D$^0$ mesons for PbPb $\sqrt{s} = 2.76$ TeV (left) and
    $\sqrt{s} = 5.02$ TeV (right) semi-central collisions as function of
    $p_T$.  Gray area indicates the $p_T$ region where effects of coalescence
    may be significant.  Experimental data from Refs.~\cite{Abelev:2014ipa,
    CMS:2016jtu}.}
  \label{fig:cumulants}
\end{figure}

We show in Fig.~\ref{fig:raa_meson} a comparison of \Dmeson \raa computed by
\dabmod with experimental data~\cite{Adam:2015sza, CMS-PAS-HIN-15-005,
CMS-PAS-HIN-16-001} for both $\sqrt{s} = 2.76$ TeV and $\sqrt{s} = 5.02$ TeV
PbPb central collisions.  In the considered region of $p_T \gtrsim 10$ GeV
our results lead to a good agreement with the data and are similar for both
fragmentation temperatures of $T_d = 120$ MeV and $T_d = 160$ MeV.

In Fig.~\ref{fig:cumulants} we compute the multi-particle cumulants
$v_2\{m\}$ of D$^0$ mesons for the same collision energies at a different
centrality range of 30--40\% and compare with currently available
experimental data~\cite{Abelev:2014ipa, CMS:2016jtu}.  One can see that our
results are consistent with data at high $p_T$ for the two collision
energies.  At low $p_T \lesssim 10$ GeV one must consider that coalescence is
not negligible and our results fall bellow experimental data.  Furthermore,
different energy loss mechanisms come into play in the low $p_T$
regime~\cite{Nahrgang:2014vza}, which may contribute to the overall magnitude
of the computed flow harmonics.

\begin{figure}[h!]
  \centering
  \includegraphics{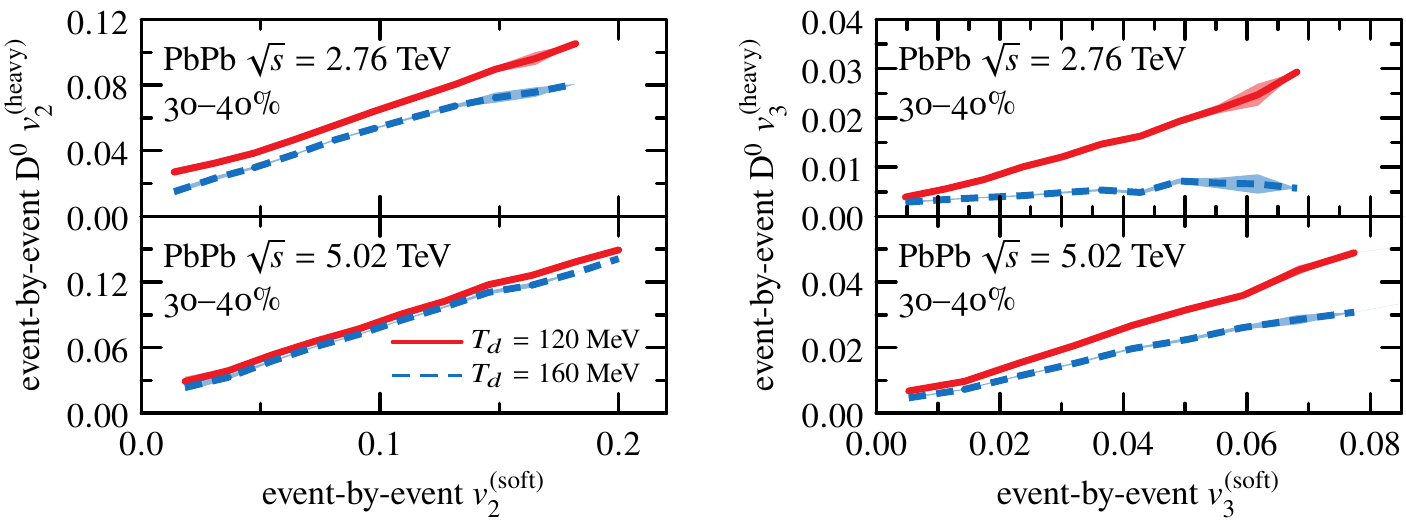}
  \caption{(Color online) Flow coefficients $v_n$ correlations between  the
    heavy sector for \Dmeson and the soft sector for $n=2$ (left) and $n=3$
    (right) for different temperatures for the fragmentation $T_d$.}
  \label{fig:correlation}
\end{figure}

Using event-by-event correlations~\cite{Prado:2016szr} one can examine
different parameters of the simulation and study their effects. In
Fig.~\ref{fig:correlation} we show the correlations for PbPb at $\sqrt{s} =
2.76$ TeV and $\sqrt{s} = 5.02$ TeV semi-central collisions.  The left
(right) plots show the correlation for the elliptic (triangular) flow. The
figure exhibits a clear difference of the slopes for the two chosen
fragmentation temperatures in the case of $v_3$, which is not observed for
$v_2$. That might be related to the build up time of each harmonic, since,
the higher the $T_d$, the less time the quark has to interact with the medium
before hadronization occurs. The plots indicate that for this energy loss
parametrization, $v_3$ takes longer to build up than $v_2$, which should get
most of its effect from the initial interaction with the medium.

\section{Conclusions}\label{sec:concl}

This work combines event-by-event hydrodynamic flow and temperature profiles
with a parametrization for heavy quark energy loss, which allows for the
computation of \raa and $v_2$ of \Dmeson mesons at high $p_T$.  By
implementing these calculations into a Monte Carlo simulation, called
\dabmod, we were able to obtain the correlations between the heavy flavor and
soft sectors for the elliptic and triangular flow harmonics $v_2$ and $v_3$
using an event engineering technique first described in~\cite{Prado:2016szr}.
Our results show that the $v_3$ magnitude is highly affected by the
fragmentation temperature which indicates that it might be built up at later
stages during the evolution of heavy quarks within the medium when compared
to $v_2$.

\section*{Acknowledgements}

We thank M.~Luzum for discussions and \textsc{fapesp} and \textsc{cnp}q for
support. J.N.H. was supported by NSF grant no. PHY-1513864 and she
acknowledges the use of the Maxwell Cluster and the advanced support from the
Center of Advanced Computing and Data Systems at the University of Houston to
carry out the research presented here. 

\bibliographystyle{elsarticle-num}

\end{document}